\newif\iffigs\figstrue
\documentclass[a4paper,11pt]{article}
\usepackage{latexsym,amssymb,amsmath,lscape,graphics,setspace}
\usepackage{graphicx}        
\usepackage{amsfonts}

\makeatletter \@addtoreset{equation}{section} \makeatother

\def\a{\alpha} \def\b{\beta}

\usepackage{geometry}                
\usepackage{graphicx}
\usepackage{amssymb}
\begin{document}

\begin{titlepage}
\begin{flushright}
\par\end{flushright}
\vskip 1.5cm
\begin{center}
\textbf{\huge \bf Remarks on the Integral Form of \vspace{.2cm} \\  D=11 Supergravity}
\vskip 1.5cm
\vskip 1cm
\large {\bf Pietro Antonio Grassi}$^{~a,b,}$\footnote{pietro.grassi@uniupo.it} 
\vskip .5cm {
\small
\centerline{$^{(a)}$
\it Dipartimento di Scienze e Innovazione Tecnologica (DiSIT),} 
\centerline{\it Universit\`a del Piemonte Orientale, viale T.~Michel, 11, 15121 Alessandria, Italy}
\medskip
\centerline{$^{(b)}$
\it INFN, Sezione di Torino, via P.~Giuria 1, 10125 Torino, Italy}
}
\end{center}
\vskip  .5cm
\vfill{}

\begin{abstract}

We make some considerations and remarks on $D=11$ supergravity and its integral form. 
We start from the geometrical formulation 
of supergravity and by means of the integral form technique we provide a superspace action that reproduces  
(at the quadratic level) the recent formulation of supergravity in pure spinor framework. We also make some remarks on Chevalley-Eilenberg cocycles and their Hodge duals. 

\end{abstract}
\vfill{}
\vspace{1.5cm}
\end{titlepage}
\newpage\setcounter{footnote}{0}

\section{Introduction}

The integral form of supergravity has been introduced in \cite{Castellani:2014goa,Castellani:2015paa,
Castellani:2016ibp} to provide an action principle for supersymmetric theories in the 
geometric formulation \cite{book}. It has been shown that any superspace Lagrangian for the same theory can be reached by a 
suitable choice of the super embedding of the bosonic submanifold into a supermanifold or, in alternative way,  the geometric formulation provides an interpolating model between all possible superspace realisations of the same theory.  
In the case of rigid supersymmetry, it has been shown in \cite{CCG} how to choose the superembedding via the Picture Changing Operator to reproduce all possible superspace versions of the same theory. In the case of curved dynamical geometry, there are further complications that should be taken into account in order to 
provide an action principle (simple examples are discussed in \cite{Castellani:2016ibp,Catenacci:2018jjj}).  
Besides the technical construction of different possible actions, the integral form of supergravity has the conceptual strength to clarify the difference between on-shell superspace, the absence of auxiliary fields, and 
the component formulation with dynamical equations of motion. A further aim is to 
translating the geometric understanding at the quantum level for the functional approach to quantum field theories and a solid variational principle is an important basic ingredient. The present note is not intended to solve these problems, but to collect some preliminary remarks and considerations regarding the application of the integral form of supergravity to the D=11 model \cite{Cremmer:1978km,VanNieuwenhuizen:1981ae,DAuria:1982uck}. Some concerns regarding the action principle are given in sec. 1, where it is pointed out that the equations of motion cannot emerge from a naive action principle, but they have to be supplemented by additional pieces of information since, in the geometric formulation, the geometric (a.k.a. {\it rheonomic}) Lagrangian is not closed. This is a narrow bottleneck that might prevent any simple solution to the problem. In sec. 2 
we recall some basic formulas for the super volume and the geometry of curved supermanifolds. In sec. 3 
the geometric formulation \cite{book,DAuria:1982uck,Castellani:2022iib} of D=11 supergravity is recollected. In sec. 4, 
we provide a bridge between D=11 supergravity obtained from pure spinor 
formulation of the super membrane \cite{Berkovits:2002uc,cede} and the integral form of supergravity. Notice that 
only by means of the action principle we can definitely compare the two frameworks. It has been noticed 
already in the case of D=10 super Yang-Mills, see \cite{Fre:2017bwd}, the bridge between the two formalisms,  
but not in the context of dynamical supermanifolds. We hope that the present ideas might serve to understand 
better the recent developments \cite{Borsten:2023reb} in the context of supergravity. 
Finally, in sec. 6, we describe an application of the Hodge duality on supermanifolds 
\cite{Castellani:2015ata,CCGir} on the D=11 cocycles which might provide a further tool to study the hidden symmetries as advocated in \cite{DAuria:1982uck} and in the recent works \cite{Ravera:2021sly,Andrianopoli:2017itj}.

\section{Few Remarks on Supergravity Action with PCO}

Given the $(11|0)$-superform Lagrangian $\mathcal{L}^{(11|0)} 
\in \Omega^{(11|0)} \left( \mathcal{M}^{(11|32)} \right)$, the corresponding action 
on the entire supermanifold $\mathcal{M}^{(11|32)}$ is obtained in \cite{book} 
by choosing an embedding $i: 
\mathcal{M}^{(11)} \to \mathcal{M}^{(11|32)}$ 
and defining the integral 
\begin{equation}\label{acA}
	S = \int_{\mathcal{M}^{(11)} \hookrightarrow \mathcal{M}^{(11|32)}} i^* \mathcal{L}^{(11|0)}.
\end{equation}
where $i^* \mathcal{L}^{(11|0)}$ is the pull-back on $\mathcal{M}^{(11)}$ of the full-superspace 
Lagrangian $\mathcal{L}^{(11|0)}$. 

In the framework of integral forms 
we lift the Lagrangian to a top form $\mathcal{M}^{(11|32)}$ by means of a \emph{Picture Changing Operator} (PCO) $\mathbb{Y}^{(0|32)}$ which is the \emph{Poincar\'e dual} of the embedding
 and it can be realized as a multiplicative operator on the space of form 
 $\Omega^{(p|q)} \left( \mathcal{M}^{(11|32)} \right)$. Using the usual technique in differential geometry 
 we can rewrite \eqref{acA} as
 \begin{equation}\label{acB}
    S = \int_{\mathcal{M}^{(11|32)}} \mathcal{L}^{(11|0)} \wedge \mathbb{Y}^{(0|32)}
\end{equation}
where $\mathcal{L}^{(11|0)}$ is the geometric Lagrangian and $\mathbb{Y}^{(0|32)}$ 
depends upon the embedding. 
For example, associated to the trivial embedding, we have
\begin{equation}\label{acC}
    \mathbb{Y}^{(0|32)}_{\rm{s.t.}} = \theta^1 \ldots \theta^{32} \delta \left( d \theta^1 \right) \wedge \ldots \wedge \delta \left( d \theta^{32} \right) \ ,
\end{equation}
and the corresponding action  $ S = \int_{\mathcal{M}^{(11|32)}} \mathcal{L}^{(11|0)} \wedge \mathbb{Y}^{(0|32)}_{\rm{s.t.}} = \int_{\mathcal{M}^{(11)}} \mathcal{L}^{\rm{s.t.}}$ 
where $\mathcal{L}^{\rm{s.t.}}$ is the space-time Lagrangian. 
$\mathbb{Y}^{(0|32)}$ is an element of the cohomology $H^{(0|32)} \left( \mathcal{M}^{(11|32)} , d \right)$. Changing the representative corresponds to the choice of different embeddings of the bosonic submanifold and it changes by  $d$-exact terms: $\mathbb{Y}^{(0|16)} \mapsto 
\mathbb{Y}^{(0|32)} + d \Sigma^{(-1|32)}$ where we consider negative-degree integral forms 
(see, e.g., Appendix). 

In general, the action will be independent of the choice of a representative if {$\mathcal{L}^{(11|0)}$} is closed:
\begin{align}
    \nonumber S' &= \int_{\mathcal{M}^{(11|32)}} \mathcal{L}^{(11|0)} \wedge \mathbb{Y}'^{(0|32)} = \int_{\mathcal{M}^{(11|32)}} \mathcal{L}^{(11|0)} \wedge \left( \mathbb{Y}^{(0|32)} + d \Sigma^{(-1|32)} \right) \\
    \label{acD} &=\int_{\mathcal{M}^{(11|32)}} \mathcal{L}^{(11|0)} \wedge \mathbb{Y}^{(0|32)} 
    - \int_{\mathcal{M}^{(11|32)}} d \mathcal{L}^{11|0)} \wedge \Sigma^{(-1|32)} + \rm{b.t.} \nonumber\\
    &= S - \int_{\mathcal{M}^{11|32)}} d \mathcal{L}^{(11|0)} \wedge \Sigma^{(-1|32)} + \rm{b.t.} \ ,
\end{align}
where with ``b.t.'' we denote boundary terms. However, the closure of the Lagrangian is guaranteed only in a few known cases, in particular, when it is possible to add auxiliary fields that guarantee off-shell invariance of the Lagrangian. For $D=11$ supergravity, $d  \mathcal{L}^{(11|0)} \neq 0$ and therefore we are not authorised to change the PCO $\mathbb{Y}'^{(0|32)}$. 

The  Euler-Lagrange equations derived from the Lagrangian $\mathcal{L}^{(11|0)}$ 
do not coincide with the equations of motion coming from a variational principle of an action, as they have 
of take into account the PCO  $\mathbb{Y}^{(0|32)}$. Varying with respect to any field $\phi$ of the theory we get
\begin{equation}\label{actD}
	\delta_\phi S = 0 \ \implies \ 
	\int_{\mathcal{M}^{(11|32)}} \delta_\phi \mathcal{L}^{(11|0)} \left( \phi \right) \wedge \mathbb{Y}^{(0|32)} + 
	\mathcal{L}^{(11|0)} \left( \phi \right) \wedge  \delta_\phi  \mathbb{Y}^{(0|32)}= 0 \ .
\end{equation}
Then, by using the requirement that $ \mathbb{Y}^{(0|32)}$ is a representative of the cohomology and if  
the variation $ \delta_\phi  \mathbb{Y}^{(0|32)}$ can be expressed as a Lie derivative (for example by changing the embedding of the sub manifold into the supermanifold), it follows $\delta_\phi  \mathbb{Y}^{(0|32)} = d  \mathbb{J}^{(-1|32)}$. Thus, by integration by parts and neglecting boundary terms,  we get 
\begin{eqnarray}
\label{actE}
\delta_\phi \mathcal{L}^{(11|0)} \left( \phi \right) \wedge \mathbb{Y}^{(0|32)} + 
	d \mathcal{L}^{(11|0)}  \wedge  \mathbb{J}^{(-1|32)}= 0
\end{eqnarray}
Different choices of PCO reflect different superspace Lagrangian with different amount of manifest isometries when $d \mathcal{L}^{(11|0)}\neq 0$ 
and the superspace Euler-Lagrangian equations $\delta_\phi \mathcal{L}^{(11|0)} =0$, 
as discussed in \cite{book}, can only be derived from $S$ 
if $d \mathcal{L}^{(11|0)} =0$ and if $\mathbb{Y}^{(0|32)}$ has no kernel. 
For a generic variation, we have to take into account also the variation of the PCO and the true equations 
of motion are 
\begin{eqnarray}
\label{actF}
\delta_\phi \mathcal{L}^{(11|0)} \left( \phi \right) \wedge \mathbb{Y}^{(0|32)} + 
	\mathcal{L}^{(11|0)} \left( \phi \right) \wedge  \delta_\phi  \mathbb{Y}^{(0|32)} =0
\end{eqnarray}
If the PCO $ \mathbb{Y}^{(0|32)}$ is independent from any supergravity field $\phi$, 
the second term drops and we are left with the equations 
\begin{eqnarray}
\label{actG}
\delta_\phi \mathcal{L}^{(11|0)} \left( \phi \right) \wedge \mathbb{Y}^{(0|32)} =0
\end{eqnarray}
which are not equivalent to $\delta_\phi \mathcal{L}^{(11|0)} \left( \phi \right) =0$ unless $ \mathbb{Y}^{(0|32)}$ 
has no kernel. Let us consider the choice \eqref{acC}. It is evident that any form $\omega(\theta, d\theta)$ 
which is at least linear in $\theta$ and in $d\theta$ is in its kernel. However, there is another representative 
\begin{eqnarray}
\label{actH}
 \mathbb{Y}^{(0|32)}_{no-\theta-ker} =  (1 +  \prod_{\alpha=1}^{32} \theta^{\alpha})  \prod_{\alpha=1}^{32} \delta(d\theta^{\alpha}) = \mathbb{Y}^{(0|32)}_{s.t.} + d \left( -\frac{1}{32}\theta^{\beta} \iota_{\beta}  \prod_{\alpha=1}^{32} \delta(d\theta^{\alpha})\right)  
\end{eqnarray}
where $\iota_{\alpha} = \partial/\partial_{d\theta^{\alpha}}$ is the contraction along the odd vector field 
$\partial_{\alpha}$. $\mathbb{Y}^{(0|32)}_{no-\theta-ker}$ has a smaller kernel since only linear functions 
in $d\theta$ are in its kernel. In order to provide a no-kernel PCO, we should add further contractions $\iota_{\alpha_{1}} \dots \iota_{\alpha_{k}} $ 
along $k$ odd vector fields, but that decreases the form number to $-k$. That can be compensated 
by factors of forms $dx^{a_{1}} \wedge \dots \wedge dx^{a_{k}}$ which have a non trivial kernel.  
Therefore, it appears rather difficult to have a field-independent PCO $ \mathbb{Y}^{(0|32)}$ which has 
no kernel to justify the equations $\delta_\phi \mathcal{L}^{(11|0)} \left( \phi \right) =0$. 
Of course, the latter implies \eqref{actG}, but not vice-versa. The study of the kernel of 
$ \mathbb{Y}^{(0|32)}$ is anyway important to understand the supergravity equations of motion. 

There is a further possibility \cite{Berkovits:2009gi,Berkovits:2006bk,Berkovits:2006vi}. 
The PCO \eqref{acC} can be rewritten as (up to a unessential coefficient $\#$) 
\begin{eqnarray}
\label{actI}
\mathbb{Y}^{(0|32)} = \# \int [d^{32}{w} d^{32}p] e^{i p_{\alpha} \theta^{\alpha} + i w_{\alpha} d \theta^{\alpha}}
\end{eqnarray}
where we add some auxiliary variables $w_{\alpha}$ and $p_{\alpha}$, which are commuting and anticommuting, respectively. Integrating on the latter we retrieve the original PCO \eqref{acC}, but 
before integrating this expression has no kernel and it might serve to define the superspace equations 
of motion. 
Checking that $d \mathbb{Y}^{(0|32)}=0$ is easy. However, the introduction of the new set of coordinates 
$w_{\alpha}$ and $p_{\alpha}$ might introduce new degrees of freedom since the fields $\phi$ and 
the Lagrangian could depend upon them. To avoid this, in \cite{Berkovits:2009gi,Berkovits:2006bk,Berkovits:2006vi} the differential is changed into 
$ d  \mapsto d + p_{\alpha} \partial_{w_{\alpha}}$ such that $d w_{\alpha} = p_{\alpha}$ and $d p_{\alpha}=0$, 
which is equivalent to say that the $d$-cohomology does not depend on them.  Therefore, the Lagrangian would be independent of them if it would be an element of the cohomology, namely $d \mathcal{L}^{{11|0}} =0$ 
and $\mathcal{L}^{{11|0}} \neq d \Sigma$, but that it is not the case and therefore also this path is obstructed.

\section{Super-Vielbeins and the Super-Volume Form}

We consider $D=11$ $N=1$ superspace parametrized by the coordinates $(x^m, \theta^\mu)$ with 
$m=0,\dots,10$ and $\mu=1,\dots,32$. We use the Majorana representation for the spinors $\theta^\mu$ using 
real coordinates. We use $a=0,\dots,10$ and $\alpha =1,\dots,32$, for flat indices of the target space with  
the supercharges defined as
 $Q_\a = \partial_\a + (\Gamma^a \theta)_\alpha \partial_a$ and $P_a = \partial_a$. 
The $\Gamma^a$ are 11-d Dirac matrices with
$\{\Gamma^a, \Gamma^b\} = 2 \eta^{ab}$. $\eta^{ab}$ is the flat metric on the tangent space, the signature 
is $(+,-, \dots,-)$. The relevant Fierz identities are
$(\bar\psi \Gamma^{ab} \psi) (\bar\psi \Gamma_a \psi) =0$   
and the bi-spinor decomposition is 
\begin{eqnarray}
\label{curD}
\psi {}_\wedge \bar \psi = 
\frac{1}{32} \left( \Gamma_a (\bar \psi \Gamma^a \psi) - \frac12 \Gamma_{a b} 
(\bar \psi \Gamma^{ab} \psi) + \frac{1}{5!}  
\Gamma_{a_1 \dots a_5} (\bar \psi \Gamma^{a_1 \dots a_5} \psi) \right) 
\end{eqnarray}

The supervielbeins $V^a$ and $\psi^\alpha$ are $(1|0)$-superforms with values in the tangent space 
and they can be decomposed on the basis $(dx^m, d\theta^\alpha)$ as
\begin{eqnarray}
\label{exC}
V^a = E^a_m \, dx^m + E^a_\mu d\theta^\mu\,, ~~~~~~~~~~~
\psi^\alpha = E^\alpha_m \, dx^m + E^\alpha_\mu d\theta^\mu\,, 
\end{eqnarray}
We can write
\begin{eqnarray}
\label{exCA}
T^a =\mathcal{D} V^a - \bar\psi^\alpha \Gamma^a_{\alpha\beta} \psi^\beta\,, ~~~~~~
\rho^\alpha = \mathcal{D}  \psi^\alpha\,. 
\end{eqnarray}
with $T^a, \rho^\a$ the vectorial and spinorial parts of the super torsion, respectively. 
The covariant derivatives are defined as $\mathcal{D}  V^a = d V^a + \varpi^a_{~b} V^b$ and $\mathcal{D} \psi^\alpha = d \psi^\alpha + \frac14 \varpi_{ab} 
\Gamma^{ab, \alpha}_{\beta} \psi^\beta$ where $\varpi^{ab}$ is the spin connection. The equations 
$\varpi^\alpha_{~\beta} = 
\frac14 (\Gamma^{ab})^\alpha_{~\beta} \varpi_{ab}$ relate the spinorial representation with the vector representation. 
If we set $T^a=\rho^{\alpha} =0$ in order to fix the spin connection $\varpi_{ab}$ 
in terms of $(V^a, \psi^\a)$. In general, for non-vanishing dynamical supergravity fields, we cannot also set $\rho^\a =0$. The supermatrix 
\begin{eqnarray}
\label{exD}
\mathbb{E} = \left(
\begin{array}{cc}
 E^a_m  &   E^a_\mu    \\
  E^\alpha_m  &   E^\alpha_\mu 
\end{array}
\right)
\end{eqnarray}
is the well-known supervielbein appearing in supegravity. In the case of flat space, it reads 
\begin{eqnarray}
\label{exDA}
\mathbb{E}_{flat}\equiv 
(V^a, \psi^\alpha) = (dx^{a}+ \bar\theta \Gamma^{a} d\theta,d\theta^{\alpha})\ ,
\end{eqnarray}
and the fluctuations around $\mathbb{E}_{flat}$ in \eqref{exDA} are identified with the dynamical vielbein and gravitino. 
Given a supermatrix, we can define the invariant superfields  
\begin{eqnarray}
\label{exE}
{\rm Ber}(\mathbb{E}) = \frac{\det(E^a_m - E^a_\mu (E^{-1})^\mu_{~\alpha} E^\alpha_m)}{\det E^\alpha_\mu} = 
 \frac{\det E^a_m }{\det (E^\alpha_\mu  - E^\alpha_m (E^{-1})^m_{~a} E^a_\mu )} \ ,
\end{eqnarray}
which are the two equivalent expressions of the {super determinant}, well-defined when $\det E^a_m \neq 0$ and 
$\det E^\a_\mu \neq 0$. 

The volume form is the integral form with form number $11$ and picture number $32$ (see the appendix for an introduction to integral forms and related notations):
\begin{eqnarray}
\label{exF}
{\rm {Vol}}^{(11|32)} = \epsilon^{a_1 \dots a_{11} } \delta(V^{a_1})_\wedge \dots {}_\wedge\delta(V^{a_{11}}) \epsilon^{\alpha_1 \dots \alpha_{32}} \delta(\psi^{\alpha_1})_\wedge \dots {}_\wedge\delta(\psi^{\alpha_{32}})   \ ,
\end{eqnarray}
which corresponds to a top form in supergeometry. 
It is closed, $d{\rm Vol}^{(11|32)}= \nabla {\rm Vol}^{(11|32)} = 0$, and the non-exactness depends on the $(11|32)$ supermanifold on which it is defined (for example, if the supermanifold is compact, the top form is non-exact). The cohomological properties can be easily checked by applying the differential $d$, the Leibniz rule for $\nabla$ acting on the supervielbeins $V^a, \psi^\alpha$ and the 
distributional identity $\psi^\a \delta(\psi^\a)=0$.

In addition, the volume form ${\rm Vol}^{(11|32)}$ is invariant with respect to Lorentz transformations  
\begin{eqnarray}
\label{exFA}
\delta V^a = \Lambda^a_{~b} V^b\,, ~~~~~~~~~
\delta \psi^\alpha = \frac14 \Lambda_{ab} (\Gamma^{ab})^\alpha_{~\beta} \psi^\beta \ ,
\end{eqnarray}
then, more precisely, it belongs to the equivariant cohomology. 

Since the $V^a$'s are anticommuting (bosonic 1-forms), we can replace the first delta's with their arguments
\begin{eqnarray}
\label{exG}
{\rm Vol}^{(11|32)} = \epsilon_{a_1 \dots a_{11}}V^{a_1}_\wedge \dots {}_\wedge V^{a_{11}} \epsilon^{\alpha_1 \dots \alpha_{32}} \delta(\psi^{\alpha_1})_\wedge \dots {}_\wedge \delta(\psi^{\alpha_{32}}) \ .
\end{eqnarray}
The same cannot be done for $\psi^\alpha$'s, since they are commuting. Notice that the $\delta(\psi^\a)$ does not transform as a tensor with respect to 
change of parametrization and therefore the index $\alpha$ is not a conventional covariant index summed with the 
Levi-Civita tensor $ \epsilon^{\alpha_1 \dots \alpha_{32}}$; the latter serves only to keep track of the order of the deltas, because of their anticommutation relations 
$\delta(\psi^\a)_\wedge \delta(\psi^\beta) = - \delta(\psi^\beta)_\wedge \delta(\psi^\a)$. Nonetheless, expression \eqref{exG} is invariant under reparametrizations. 
One has to pay some attention to deal with those ``covariant" expressions. 
 
Inserting expressions (\ref{exC}) into (\ref{exG}), 
using the properties of the oriented delta's and of the 1-forms, we get 
\begin{eqnarray}
\label{exH}
{\rm Vol}^{(11|32)}  = {\rm Ber}(\mathbb{E}) \, 
\epsilon_{m_1 \dots m_{11} }dx ^{m_1}_\wedge \dots {}_\wedge dx^{m_{11}} \epsilon^{\mu_1 \dots \mu_{32}} \delta(d\theta^{\mu_1})_\wedge \dots {}_\wedge \delta(d\theta^{\mu_{32}}) \ ,
\end{eqnarray}
where the overall factor is the superdeterminant of $\mathbb{E}$ \eqref{exF}, as expected. Finally, 
\begin{eqnarray}
\label{exI}
\int_{{SM}^{{(11|32)}}} {\rm Vol}^{(11|32)} = \int  {\rm Ber}(\mathbb{E}) [d^{11}x d^{32}\theta]
\end{eqnarray}
gives the volume of the supermanifold ${SM}^{{(11|32)}}$.

\section{D=11 Supergravity in the Geometric Framework}

In order to deal with D=11 supergravity, we refer to standard literature for the 
action and the details (see for example \cite{Cremmer:1978km,VanNieuwenhuizen:1981ae,book}), but we adopt the notations and the definitions given in \cite{book}. The physical degrees of freedom are  
$V^a, \omega^{ab}, \psi^\a, A^{(3|0)}$ the first three fields are $(1|0)$-forms the last one 
is a $(3|0)$. In addition, we add also the $(6|0)$-form $B^{(6|0)}$ by consistency. 
 
The corresponding curvature are given by 
\begin{eqnarray}
\label{curA}
R^{ab} &=& d \omega^{ab} - \omega^{ac}  \omega_{c}^{~b}\,, \nonumber \\
T^a  &=& {\cal D} V^a - \frac{i}{2} \bar \psi  \Gamma^a \psi = 
d V^a - \omega^{a}_{~b} V^b - \frac{i}{2} \bar \psi  \Gamma^a \psi \,, \nonumber \\
\rho &=& {\cal D} \psi = d \psi - \frac14 \omega_{ab}\Gamma^{ab} \psi\,, \nonumber \\
F &=& d A - \frac12 \bar \psi \Gamma_{ab} \psi V^a  V^b\,, \nonumber \\
H &=& d B - \frac{i}{2} \bar \psi \Gamma_{a_{1} \dots a_{5}} \psi V^{a_{1}} \dots V^{a_{5}} - \frac{15}{2} \bar\psi \Gamma_{ab} \psi V^{a}V^{b} \wedge A - 15 F\wedge A\,.  
\end{eqnarray}
where $R^{ab}$ is the curvature of the connection, $T^a$ is the torsion, $\rho^{\alpha}$ is the torsion of the fermionic component of the supervielbein which is also identified with the curvature of the gravitino. $F$ is the $4$-form field strength of the $3$-form and $H$ is the field strength of the $6$-form $B$. In the flat space, 
namely, when all curvatures are zero, we see that 
\begin{eqnarray}
\label{curAB}
 \omega^{(4|0)} &=&  \frac12 
 (\bar\psi \Gamma_{ab} \psi) V^a V^b\,,  ~~~~d \omega ^{(4|0)} =0\,. 
\nonumber \\
d A^{(3|0)} &=& \omega^{(4|0)}\,, ~~~~~ \nonumber \\
\omega^{(7|0)} &=& \frac{i}{2} (\bar\psi \Gamma_{a_{1} \dots a_{5}} \psi) 
V^{a_{1}} \dots V^{a_{5}}\,,  ~~~~d \omega ^{(7|0)} = \omega^{(4|0)} \wedge \omega^{(4|0)}\,. 
\nonumber \\
 d B^{(6|0)} &=& \omega^{(7|0)} + 15 \omega^{(4|0)} \wedge A 
\end{eqnarray}
where $ \omega^{(4|0)}$ and $\omega^{(7|0)} + 15 \omega^{(7|0)} \wedge A$ are 
two Chevalley-Eilenberg Cohomology class for the differential $d$ in the case of flat space and 
the field $A^{(3|0)}$ and $B^{(6|0)}$ are the potentials for the free differential algebra. 

Acting with the differential on the curvatures, one obtains the Bianchi identities
\begin{eqnarray}
\label{curE}
&& {\cal D} R^{ab} =0\,,  \\
&& {\cal D} T^a + R^{a}_{~b} V^b - i (\bar\psi \Gamma^a \rho) =0\,, \nonumber \\
&& {\cal D} \rho + \frac14 \Gamma_{ab} \psi  R^{ab} =0\,, \nonumber \\
&& d F - (\bar \psi  \Gamma_{a b} \rho)  V^{a} V^{b} + 
(\bar \psi  \Gamma_{ab} \psi)
T^{a} V^{b}=0\,, \nonumber \\
&& d H - i (\bar\psi \Gamma_{a_{1} \dots a_{5}} \rho) V^{a_{1}} \dots V^{a_{5}} - 
\frac{5i}{2}(\bar \psi \Gamma_{a_{1} \dots a_{5}} \psi) T^{a_{1}} \dots V^{a_{5}} - 
15 (\bar\psi \Gamma_{ab} \psi) V^{a} V^{b}\wedge F - 15 F\wedge F \,, \nonumber 
\end{eqnarray}
which relate the curvature and the supervielbeins $(V^{a}, \psi^{\alpha})$.

Using the constraints 
\begin{eqnarray}
\label{curEA}
T^{a} &=&0\,, \nonumber \\
F  &=& F_{a_{1} \dots a_{4}} V^{a_{1}} \dots V^{a_{4}}\,, \nonumber \\
H &=& H_{a_{1} \dots a_{7}} V^{a_{1}} \dots V^{a_{7}}\,, \nonumber \\
\rho^{\alpha} &=& \rho^{\alpha}_{ab} V^{a} V^{b}\,, \nonumber \\
R^{ab} &=& R^{ab}_{cd}  V^{c} V^{d} + \bar\Sigma^{ab}_{\alpha c} \psi^{\alpha} V^{c} + 
\psi^{\alpha} K^{ab}_{\alpha\beta} \psi^{\beta} \,. 
\end{eqnarray}
where $ F_{a_{1} \dots a_{4}}, H_{a_{1} \dots a_{7}},  \rho^{\alpha}_{ab} , 
R^{ab}_{cd} , \bar\Sigma^{ab}_{\alpha c}, K^{ab}_{\alpha\beta}$ are unconstrained 
superfields. These constraints are very strong and together with the Bianchi identities,
they imply the equations of motion for the field components of \eqref{curEA}. 
The superfields $\bar\Sigma^{ab}_{\alpha c}, K^{ab}_{\alpha\beta}$ are fixed in terms 
of the other superfields. 

In \cite{book} the rheonomic Lagrangian is provided and it reads
\begin{eqnarray}\label{lagA}
{\cal L}^{(11|0)} 
&=& 
\frac{1}{330} F_{a_1 \dots a_4} F^{a_1 \dots a_4} V^{c_1} \dots {}_\wedge 
V^{c_{11}} \epsilon_{c_1 \dots c_{11}} \nonumber \\
&-& {\frac19} R^{a_1 a_2} {}_\wedge V^{a_3} \dots {}_\wedge V^{a_{11}} \epsilon_{a_1 \dots a_{11}} \nonumber \\
&+& 2 (\bar \rho {}_\wedge \Gamma_{c_1 \dots c_8} \psi) {}_\wedge V^{c_1} \dots {}_\wedge V^{c_8} \nonumber \\
&+&
 \left( \frac14 (\bar\psi {}_\wedge \Gamma^{a_1 a_2} \psi) {}_\wedge 
(\bar\psi {}_\wedge \Gamma^{a_3 a_4} \psi) +  2 F  F^{a_1 \dots a_4} \right)
{}_\wedge V^{a_5} 
\dots {}_\wedge V^{a_{11}} \epsilon_{a_1 \dots a_{11}} \nonumber \\
&+& \frac{7 i}{30} T^a {}_\wedge V_a {}_\wedge (\bar \psi {}_\wedge \Gamma^{b_1 \dots b_5} 
\psi) {}_\wedge V^{b_6} \dots {}_\wedge V^{b_{11}} \epsilon_{b_1 \dots b_{11}} \nonumber \\
&-& 84 F {}_\wedge \omega^{7}
+ 840  F {}_\wedge A {}_\wedge \omega^{4}
- 210 \,  A{}_\wedge \omega^{4} {}_\wedge \omega^{4}  - 840 F {}_\wedge F {}_\wedge A \nonumber 
\end{eqnarray}
The fourth line will be absent if we set $T^{a}=0$; this single constraint is not sufficient to put the theory on-shell. Note that the Lagrangian being a superform can be expanded into $V^{a}$ and $\psi^{\alpha}$, but we 
can select those terms which are explicitly depending on $V^{a}$ ($\rho, R^{ab}, F, H, T^{a}$ depend implicitly on $V$'s) as follows
\begin{eqnarray}
\label{lagB}
{\cal L}^{(11|0)}  = \sum_{k=0}^{11} \mathcal{L}_{a_{1} \dots a_{k}} V^{a_{1}} \dots V^{a_{k}}
\end{eqnarray}
where $ \mathcal{L}_{a_{1} \dots a_{k}} \neq 0$ if $k=0,2,4,5,6,7,8,9,11$. If the PCO has 
several factors of $V$'s, acting multiplicatively ${\cal L}^{(11|0)}  \wedge \mathbb{Y}^{(0|32)}$ 
it will kill several terms in the sum \eqref{lagB} simplifying the final superspace action. 
The choice of particular terms has to be motivated by symmetry requirements, for example 
manifest supersymmetry. In the following section, we provide an example of PCO which has been 
constructed in pure spinor supermembrane framework which select the term with $k=2$ which 
is a Chern-Simons-like term $\int A d A\wedge \omega^{4}$.

\section{PCO, Membranes and the Integral Form of the Action}

In the pioneering works \cite{Berkovits:2002uc} and later in \cite{cede} a 
supergravity action is built in the pure spinor formulation (later it has been studied from 
superparticle point of view in \cite{Anguelova:2004pg} and completed in \cite{Berkovits:2019szu,Guillen:2019pnz,Guillen:2020mmd}). Here would like to build a bridge between 
the pure spinor formulation and the geometric formulation and this can be done using the integral 
form formalism. 

The action of supergravity obtained in \cite{Berkovits:2002uc} is built in terms of a pure spinor superfield 
$C^{(3)} = \lambda^{\alpha} \lambda^{\beta} \lambda^{\gamma} C_{\alpha\beta\gamma}(x, \theta)$ where 
$\lambda$'s are the pure spinor (complex) coordinates that satisfy the quadratic constraints 
$\lambda \Gamma^{a} \lambda =0$.\footnote{Solving the constraints yields that 
23 linear independent complex pure spinors $\lambda$ and carrying a non-linear representation of the 
Lorentz group.} The components $C_{\alpha\beta\gamma}(x, \theta)$ 
are defined up to a local gauge transformation $\delta C_{\alpha\beta\gamma}(x, \theta) = 
\Gamma^{a}_{(\alpha\beta} \Sigma_{\beta) a}(x, \theta)$. The second crucial ingredient 
is the ghost-one BRST charge $Q = \int d\sigma \lambda^{\alpha}  \nabla_{\alpha}$ which is nilpotent 
if $\lambda$'s satisfy the pure spinor constraints. It has been verified that $Q C^{(3)} =0$ 
with the symmetries $\delta C^{(3)} = Q \Xi^{(2)}$ implies the linearised supergravity equations 
of motion. Then, the following action 
\begin{equation}\label{PSSA}
	S_{sugra} = \int \left[ d^{11} x d^{32} \theta D \lambda \right] C^{(3)} Q  C^{(3)}
\end{equation}
is a reasonable starting point for the complete supergravity action (see \cite{cede} for 
further developments) where the Lagrangian carries ghost number $+7$. The integral is over the 
bosonic $D=11$ coordinates, the fermionic coordinates, and over the pure spinor coordinates 
$\lambda^{\alpha}$. However, for them, we need a special measure $[D\lambda]$ 
to be compatible with the pure spinor constraints and taking into account that the pure spinors are commuting 
variables. In pure spinor cohomology, there exists the following representative at ghost number $+7$ 
\begin{equation}\label{PSSB}
 \Omega^{(7)}(\lambda, \theta) =  \epsilon_{a_1 \ldots a_{11}}  \lambda \Gamma^{a_1} \theta \ldots \lambda \Gamma^{a_7} \theta\theta \Gamma^{a_8 a_9 a_{10} a_{11}} \theta \ .
\end{equation}
which is a Lorentz singlet and explicitly depends upon 9 $\theta$'s. 
Then, the measure $[D\lambda] = d^{23}\lambda \mu(\lambda,\theta)$ is defined such that 
\begin{eqnarray}
\label{PSSC}
\int   \left[  d^{32} \theta d^{23}\lambda 
\mu(\lambda, \theta) \right] \Omega^{(7)}(\lambda, \theta) = 1
\end{eqnarray}
where $d^{32} \theta$ is the conventional Berezin integral and 
\begin{equation}\label{PSSCA}
	\mu(\lambda, \theta)  = \left( \theta^{23} \epsilon \right)_{\alpha_1 \ldots \alpha_{9}} T^{[\alpha_1 \ldots \alpha_9] (\beta_1 \ldots \beta_7 )} \frac{\partial}{\partial \lambda^{\beta_1}} \ldots \frac{\partial}{\partial \lambda^{\beta_7}} \delta^{23} \left(\lambda \right) \ ,
\end{equation}
where $T^{[\alpha_1 \ldots \alpha_9](\beta_1 \ldots \beta_7 )}  = 
\epsilon^{a_{1} \dots a_{11}}
\Gamma^{\alpha_{1} \beta_{1}}_{a_{1}} \dots \Gamma^{\alpha_{7} \beta_{7}}_{a_{7}} 
\Gamma_{a_{8} \dots a_{11}}^{\alpha_{8} \alpha_{9}}$. Notice that it depends upon the complementary 
23 $\theta$'s of $\Omega^{7}$ such that the Berezin integral gives exactly one, the derivatives ${\partial}/{\partial \lambda^{\beta}}$ act by integration-by-parts 
on $\Omega^{(7)}$ and finally the integration $\delta^{23} \left(\lambda \right) d^{23} \lambda$ gives one by 
the conventional definition of Dirac delta distributions. Finally, applying the formula \eqref{PSSC} to 
the Lagrangian in \eqref{PSSA}, one selects several pieces reproducing the quadratic part of the 
supergravity action. 

Still working on flat space, we define the new PCO 
\begin{eqnarray}
\label{PSSD}
\mathbb{Y}^{(0|32)} = 
\epsilon_{\a_1 \dots \a_{32}} \theta^{\a_1} \dots \theta^{\a_{23}} (V_{a_1} 
\Gamma^{a_1} \iota)^{\a_{24}} \dots (V_{a_9} \Gamma^{a_9} \iota)^{\a_{32}} \delta^{32}(\psi)
\end{eqnarray}
which is equivalent to the $\mathbb{Y}^{(0|32)}_{s.t.}$ since in the flat space  $V^{a} = dx^{a} + 
\bar\theta\Gamma^{a} \psi$ and by integration-by-part, we see that differs from $\mathbb{Y}^{(0|32)}_{s.t.}$ 
by exact terms. Note that the form degree is zero since the form degree of $V$'s compensate the form 
degree of $\iota_{\alpha}$. Now, we observe that if we replace $\lambda$ with $\psi$'s in 
\eqref{PSSB}, namely $\Omega^{(7)}(\lambda, \theta) \mapsto \Omega^{(7|0)}(\psi, \theta)$ we get a 7-form, but clearly it will be not closed since the $\psi$ are not pure spinors. 
Nonetheless, given the Chevalley-Eilenberg cohomology $\omega^{(4|0)}$ discussed in the 
previous section we have 
\begin{eqnarray}
\label{PSSE}
\mathbb{Y}^{(0|32)} \wedge \omega^{(4|0)} = \mu^{(-7|0)} {\rm Vol}^{(11|32)}
\end{eqnarray}
where ${\rm Vol}^{(11|32)}$ is the super volume of the manifold and $ \mu^{(-7|0)} $ 
is given by 
\begin{eqnarray}
\label{PSSF}
 \mu^{(-7|0)} = 
 \left( \theta^{23} \epsilon \right)_{\alpha_1 \ldots \alpha_{9}} T^{[\alpha_1 \ldots \alpha_9] (\beta_1 \ldots \beta_7 )} \iota_{\beta_1} \ldots \iota_{\beta_7}
\end{eqnarray}
where we replaced the derivatives w.r.t. $\lambda$ with the contraction along the odd vector fields $\nabla_{\alpha}$. The structure is exactly the same as constructed in \eqref{PSSCA} and finally 
inserting the PCO $\mathbb{Y}^{(0|32)}$ in the action, it selects the Lagrangian which is dual to 
$\Omega^{(7|0)}(\psi, \theta)$ as the pure spinor measure $\mu(\lambda, \theta)$ is dual to the 
cohomology class $\Omega^{(7)}(\lambda, \theta)$. Therefore, instead of using the pure spinor cohomology class, which we do not have in our framework, we use the Chevalley-Eilenberg cohomology $\omega^{(4|0)}$
which is well-defined and it plays a crucial role in the D=11 supergravity construction. 

Then, plugging the PCO $\mathbb{Y}^{(0|32)}$ we select only two terms. The other drop out because of the number of $V$'s at the first order. 
\begin{eqnarray}
\label{curG}
\int_{{\mathcal{SM}^{(11|32)}}} {\cal L}^{(11|0)}  \wedge \mathbb{Y}^{(0|32)} \mapsto \int_{{\mathcal{SM}^{(11|32)}}}
  \left( F \wedge A {}_\wedge (\bar\psi {}_\wedge \Gamma_{ab} \psi) {}_\wedge V^a {}_\wedge V^b 
-  F {}_\wedge F {}_\wedge A 
\right)\wedge  \mathbb{Y}^{(0|32)}
\end{eqnarray}
Notice that the result is not the complete answer, but it gives only an indication that 
the supergravity action in the geometric formulation in \eqref{lagA}
contains a superspace action similar to pure spinor formulation \eqref{PSSA}. In order to get the full 
result one needs to convert $\mathbb{Y}^{(0|32)}$ to a curved one where the $V^{a}$ and $\psi^{\alpha}$ 
in \eqref{PSSD} are the dynamical fields. This program will be tackled in subsequent publications and 
here we only discuss a first step toward the complete construction. 

A curved PCO is constructed as follows (see \cite{Cremonini:2021vyy}). We introduce the super-Euler vector as follows 
\begin{eqnarray}
\label{APPG}
&&{\mathcal E} = \theta^\mu \partial_\mu + f^m(x,\theta) \partial_m =  X^a \nabla_a + \Theta^\alpha \nabla_\a \ , \nonumber \\ 
 &&\Theta^\a = \theta^\mu E^\a_\mu + f^m(x,\theta) E^\a_m \ , ~~~~~ 
  X^a = \theta^\mu E^a_\mu + f^m(x,\theta) E^a_m \ ,
\end{eqnarray}
where the combinations $(X^a, \Theta^\a)$ are the new curved coordinates with flat indices (see \cite{Wess:1992cp}). 
We can set $X^a=0$  by 
choosing the function $f^m(x,\theta) = E^m_\alpha \Theta^\a$ yielding 
\begin{eqnarray}
\label{APPH}
\iota_{\mathcal E} \psi^\a = \Theta^\a\,, ~~~
\iota_{\mathcal E} V^a = 0\,. ~~~
\end{eqnarray}
Applying the covariant differential $\nabla$ on $\Theta^\a$, we get
\begin{eqnarray}
\label{APPK}
\nabla \Theta^\a = \nabla \iota_{\mathcal E} \psi^\a = \psi^\a - \iota_{\mathcal E} \rho^\a + \Omega^\a_{~\b} \psi^\b
~~~~\Longrightarrow ~~~~
\psi^\a = [(1 + \Omega)^{-1}]^\a_{~\beta}(\nabla \Theta^\b  +  \iota_{\mathcal E} \rho^\b) \ ,
\end{eqnarray}
where $\rho^\a$ is the spinorial component of the supertorsion (field strength of the gravitino) and $\Omega^\a_{~\b}= \iota_{\mathcal{E}} \varpi_{ab} 
(\Gamma^{ab})^\a_{~\b}$ is a gauge parameter built in terms of the spin connection $\varpi$.
In general, $\iota_{\mathcal E} \rho^\a$ does not 
vanish for dynamical supergravity fields and $\Omega^\a_{~\b} = 
 \varpi_{ab, \rho} \iota_{\mathcal E} E^\rho (\Gamma^{ab})^\a_{~\b}$.  Notice that $\Omega^\a_{~\b}$ is proportional to $\Theta$, hence it could be dropped in the expression of the curved PCO because of the product of the four $\Theta$'s in front of the delta's.
By using the expression in \eqref{APPK}, the curved PCO reads
\begin{eqnarray}
\label{APPKA}
{\mathbb Y}^{(0|32)}_{curved} = \prod_{\alpha=1}^{32} \Theta^{\alpha} \delta(\nabla \Theta^{\alpha})  =\prod_{\a=1}^{32} \iota_{\mathcal E} \psi^\a 
 \delta\Big( (1 + \Omega)^\a_{~\beta} \psi^\beta - \iota_{\mathcal E} T^\a\Big) \ ;
\end{eqnarray}
it is closed because $\nabla^2 \Theta^\a = R_{ab} (\Gamma^{ab})^\a_{~\beta} \Theta^\b$, 
since the indices on the $\Theta$'s are the flat Lorentz indices 
and because of the product of all $\Theta^\a$ in front of the delta's. Moreover, since $\iota_{\mathcal E} T^\a = \Theta^\b E^b T_{\b b}^{~~\a}$, i.e., it is proportional to $\Theta$,
${\mathbb Y}^{(0|11)}$ 
can be reduced to 
\begin{eqnarray}
\label{APPKAB}
{\mathbb Y}^{(0|32)} = \prod_{\a=1}^{32} \Theta^\a \delta(\nabla \Theta^\a)  =\prod_{\a=1}^{32} \Theta^\a 
 \delta( E^\a) \ .
\end{eqnarray}
As a last remark, the PCO \eqref{APPKA} is not 
manifestly supersymmetric. As for the flat case, we can build the new PCO by introducing some 
$V$'s in the game as follows 
\begin{eqnarray}
\label{PSSDA}
\mathbb{Y}^{(0|32)} = 
\epsilon_{\a_1 \dots \a_{32}} \Theta^{\a_1} \dots \Theta^{\a_{23}} (V_{a_1} 
\Gamma^{a_1} \iota)^{\a_{24}} \dots (V_{a_9} \Gamma^{a_9} \iota)^{\a_{32}} \delta^{32}(\psi)
\end{eqnarray}
with the constraint that $T^{a}=0$ must be imposed from the beginning or as a condition 
for the closure of the PCO (see also \cite{Castellani:2016ibp} for a complete discussion in D=3 N=1 
supergravity). 

\section{D=11 Cocycles and Hodge Duality}

In this last section, we make some considerations on the 
Chevalley-Eilenberg cocycles $\omega^{(4|0)}$ and $\omega^{{(7|0)}}$  
written in terms of the gravitinos $\psi^\alpha$ and vielbeins $V^a$.  In 
particular we construct the Laplace-Beltrami operator and we act on those cocycles 
to check the Hodge theory (a complete discussion will be provided in a separate publication 
\cite{HT}, a discussion in the context of superLie algebras can also be found in \cite{Catenacci:2020ybi}). 

The cocycles  $\omega^{(4|0)}$ and $\omega^{{(7|0)}}$ satisfy the following equations 
\begin{eqnarray}
    \label{cicB}
    d \omega_4 =0\,, ~~~~~
    d \omega_7 = -\frac12 \omega_4 \wedge \omega_4\,.
\end{eqnarray}
The second equation is a consequence of the Fierz identities 
$$\bar\psi \Gamma_{a_1 \dots a_5} \psi \bar\psi \Gamma^{a_5} \psi = 
\bar\psi \Gamma_{[a_1 a_2} \psi \bar\psi \Gamma_{a_3 a_4]} \psi$$ 
Let us consider the super Hodge dual (defined in \cite{Castellani:2015ata,CCGir}) of those superforms
\begin{eqnarray}
    \label{cicC}
    \star \omega_4 &=& V^{a_1} \dots V^{a_9} \epsilon_{a_1\dots a_9 b_1b_2}\,  \bar\iota \Gamma^{b_1 b_2}\iota \delta^{32}(\psi)\,, \nonumber \\
    \star \omega_7 &=& V^{a_1} \dots V^{a_6} \epsilon_{a_1\dots a_6 b_1 \dot b_5}\, \bar\iota \Gamma^{b_1 \dots b_5}\iota \delta^{32}(\psi)\,, ~~~~
\end{eqnarray}
where $\bar\iota \Gamma^{b_1 b_2}\iota = 
\frac{\delta}{\delta \bar{\psi}}\Gamma^{b_1 b_2}\frac{\delta}{\delta{\psi}} $ are the derivatives with respect to the argument of the delta functions. 
Therefore they act by integration by parts. In particular, if we compute the wedge product of $\omega_4$ with $\star\omega_4$ (and analogously for $\omega_7$) 
we get the volume form
\begin{eqnarray}
\label{cicCA}
\omega_4 \wedge \star \omega_4 = V_1 \dots V_{11} \delta(\psi_1) \dots \delta(\psi_{32})\,, ~~~~
\omega_7 \wedge \star \omega_7 = V_1 \dots V_{11} \delta(\psi_1) \dots \delta(\psi_{32})\,. 
\end{eqnarray}

Notice that the first one has degrees $(7|32)$ (due to the presence of $9$ vielbeins and two derivatives), while the second one has degree $(4|32)$. Notice that both are closed 
\begin{eqnarray}
\label{cicD}
d \star \omega_4 &=& 9\,  (\bar\psi \Gamma^{a_1} \psi)  
V^{a_2} \dots V^{a_9} \epsilon_{a_1\dots a_9 b_1b_2}\,  \bar\iota \Gamma^{b_1 b_2}\iota \delta^{32}(\psi) \nonumber \\
&=& 
9 \, {\rm tr}(\Gamma^{a_1} \Gamma^{b_1 b_2})  
V^{a_2} \dots V^{a_9} \epsilon_{a_1\dots a_9 b_1b_2}\, \delta^{32}(\psi) =0\,, \nonumber \\
d \star \omega_7 &=&  6\, 
V^{a_2} \dots V^{a_6} \epsilon_{a_1\dots a_6 b_1 \dots b_5}\, \bar\iota \Gamma^{b_1 \dots b_5}\iota \delta^{32}(\psi) \nonumber \\
&=& 6 \, {\rm tr}(\Gamma^{a_1} \Gamma^{b_1 \dots b_5})  
V^{a_2} \dots V^{a_9} \epsilon_{a_1\dots a_6 b_1 \dots b_5}\, \delta^{32}(\psi) = 0
\end{eqnarray}
they vanish because of the trace between the gamma matrices. On the other hand, 
if we compute the Hodge dual of $d\omega_7$, we get 
\begin{eqnarray}
\label{cicE}
\star d \omega_7 = - \frac14 V^{a_1} \dots V^{a_7} \epsilon_{a_1 \dots a_7 b_1 \dots b_4}  
\bar\iota \Gamma^{b_1 b_2}\iota  \bar\iota \Gamma^{b_3 b_4}\iota \delta^{32}(\psi) 
\end{eqnarray}
Using again the Fierz identities, we can recast the derivatives 
as follows
\begin{eqnarray}
\label{cicEA}
\star d \omega_7 = - \frac14 V^{a_1} \dots V^{a_7} \epsilon_{a_1 \dots a_7 b_1 \dots b_4}  
\bar\iota \Gamma^{b_1 b_2 b_3 b_4 b_5}\iota  \bar\iota \Gamma_{b_5}\iota \delta^{32}(\psi) 
\end{eqnarray}
and then we can compute the differential
\begin{eqnarray}
\label{cicF}
d \star d \omega_7 = - \frac74 \bar\psi \Gamma^{a_1} \psi \dots V^{a_7} \epsilon_{a_1 \dots a_7 b_1 \dots b_4}  
\bar\iota \Gamma^{b_1 b_2 b_3 b_4 b_5}\iota  \bar\iota \Gamma_{b_5}\iota \delta^{32}(\psi) 
\end{eqnarray}
The integration by parts of produces two different structures, one vanishes because of the usual trace of gamma 
matrices, but the second structure gives the 
expression 
\begin{eqnarray}
\label{cicG}
d \star d \omega_7  = - \frac72 \star \omega_7 ~~~~~ \Longrightarrow ~~~~\star d \star d \omega_7 = - \frac72 \omega_7\, 
\end{eqnarray}
then finally it leads (together the vanishing of $d \star \omega_7 =0$), 
to the Laplace-Beltrami differential $\Delta = d \star d \star + \star d \star d$ acting on 
those cocycles and it yields 
\begin{eqnarray}
\label{cicH}
\Delta \omega_7 = - \frac72 \omega_7\,,  ~~~~~~ 
\Delta \omega_4 = 0\,. 
\end{eqnarray}
The second equation follows from $d \omega_4=0$. The second equation is one side of the 
Hodge theorem, since $\omega^{(4)}$ is a Chevalier-Eilenberg cocycle, $\Delta \omega_4 = 0$ 
implies that it is also the harmonic representative. Vice-versa, $\omega^{(7)}$ is not a cohomology 
class and $\Delta \omega_7 \neq 0$.

\section*{Acknowledgements}

We would like to thank C.A. Cremonini, R. Catenacci, 
L. Castellani, R. D'auria, M. Trigiante, L. Ravera, R. Norris, L. Andrianopoli, for discussions and comments. The work is partially fund by University of Eastern Piedmont with FAR-2019 projects. 
We would also like to thank O. Corrandini and V. Vitagliano for the invitation at the conference 
``Avenues of Quantum Field Theory in Curved Spacetime", held in Genoa (IT), September 2022. 

\section*{Appendix: A Brief Review on Integral Forms}

In this appendix, we want to recall the main definitions and computation techniques for integral forms. For a more exhaustive review or for a more rigorous approach to integral forms we suggest 
\cite{Witten:2012bg,Catenacci:2018xsv,Noja:2021xos}.

We consider a supermanifold ${\cal SM}^{(n|m)}$ with $n$ bosonic and $m$ fermionic dimensions. We denote the local coordinates in an open set as $(x^a, \theta^\alpha), a=1,\ldots,n , \alpha=1,\ldots,m$. A generic $(p|0)$-form, i.e., a \emph{superform}, has the following local expression
\begin{equation}\label{ABRIFA}
	\omega^{(p|0)} = \omega_{[i_1 \ldots i_r](\alpha_1 \ldots \alpha_s)} \left( x , \theta \right) dx^{i_1} \wedge \ldots \wedge dx^{i_r} \wedge d \theta^{\alpha_1} \wedge \ldots \wedge d \theta^{\alpha_s} \ , \ p=r+s \ .
\end{equation}
The coefficients $\omega_{[i_1 \ldots i_r](\alpha_1 \ldots \alpha_s)}(x,\theta)$ are a set of superfields and the indices $a_1 \dots a_r$, $\alpha_1 \dots \alpha_s$ are anti-symmetrized and symmetrised, respectively, because of the rules (we omit the ``$\wedge$" symbol)
\begin{equation}\label{ABRIFB}
	dx^i dx^j = - dx^j dx^i \ , \ d \theta^\alpha d \theta^\beta = d \theta^\beta d \theta^\alpha \ , \ dx^i d \theta^\alpha = d \theta^\alpha d x^i \ .
\end{equation}
Namely, we assign \emph{parity} $1$ to odd forms and $0$ to even forms:
\begin{equation}\label{ABRIFC}
	\left| dx \right| = 1 \ , \ \left| d \theta \right| = 0  \ .
\end{equation}
Since superforms are generated both by commuting and anti-commuting forms, we immediately see that there is no top form. 
In other words, if one looks for the analogous of the determinant bundle on a supermanifold, one has to consider a different space of forms, namely the \emph{integral forms}. A generic integral form locally reads 
\begin{equation}\label{ABRIFD}
	\omega^{(p|m)} = \omega_{[i_1 \ldots i_r]}^{(\alpha_1 \ldots \alpha_s)} \left( x , \theta \right) dx^{i_1} \wedge \ldots \wedge dx^{i_r} \wedge \iota_{\alpha_1} \ldots \iota_{\alpha_s} \delta \left( d \theta^1 \right) \wedge \ldots \wedge \delta \left( d \theta^m \right) \ ,
\end{equation}
where $\delta \left( d \theta \right)$ is a (formal) Dirac delta function and $\iota_\alpha$ denotes the interior product. The integration on $d\theta$'s is defined 
algebraically by setting 
\begin{equation}\label{ABRIFE}
	\int_{d \theta} \delta \left( d \theta \right) = 1 \,, ~~~ \int_{d \theta} f(d\theta) \delta \left( d \theta \right) = f(0)  \ ,
\end{equation}	
for a generic test function $f(d\theta)$. 	
The symbol  $\delta \left( d \theta \right)$ satisfies the usual distributional equations 
\begin{equation}\label{ABRIFEA}
	 \ \ d \theta \delta \left( d \theta \right) = 0 \,,  
	  \delta \left( \lambda d \theta \right) = \frac{1}{\lambda} \delta \left( d \theta \right) \,, 
	  d \theta  \delta^{(1)} \left( d \theta \right) = - \delta \left( d \theta \right) \,, 
	  d \theta  \delta^{(p)} \left( d \theta \right) = - p \delta^{(p-1)} \left( d \theta \right) \,, 
\end{equation}	  
We sometimes denote by $\iota_\alpha \delta(d\theta^\a) \equiv \delta^{(1)}(d\theta^\a)$. Additional properties are 
 \begin{equation}\label{ABRIFEAA}
	 \delta \Big( d \theta^\alpha \Big) \wedge \delta \left( d \theta^\beta \right) = - \delta \left( d \theta^\beta \right) \wedge \delta \Big( d \theta^\alpha \Big) \ \ , \ \ dx \wedge \delta \left( d \theta \right) = - \delta \left( d \theta \right) \wedge d x \ ,
	 \end{equation}
indicating that actually these are not conventional distributions, but rather \emph{de Rham currents}. 

Given these properties, we retrieve  a  top form among integral forms as
\begin{equation}\label{ABRIFF}
	\omega_{top}^{(n|m)} = \omega \left( x , \theta \right) dx^1 \wedge \ldots \wedge dx^n \wedge \delta \left( d \theta^1 \right) \wedge \ldots \wedge \delta \left( d \theta^m \right) \ ,
\end{equation}
where $\omega \left( x , \theta \right)$ is a superfield. The space of $(n|m)$ forms corresponds to 
the \emph{Berezinian bundle} since  the generator $dx^1 \wedge \ldots \wedge dx^n \wedge \delta \left( d \theta^1 \right) \wedge \ldots \wedge \delta \left( d \theta^m \right)$ transforms as the superdeterminant of the Jacobian.

One can also consider a third class of forms, with non-maximal and non-zero number of delta's: the \emph{pseudoforms}. A general pseudoform with $q$ Dirac delta's is locally given by
\begin{eqnarray}\label{ABRIFG}
	\omega^{(p|q)} &=& \omega_{[a_1 \ldots a_r](\alpha_1 \ldots \alpha_s)[\beta_1 \ldots \beta_q]} \left( x , \theta \right)  \\ 
	&&
	dx^{a_1} \wedge \ldots \wedge dx^{a_r} \wedge d \theta^{\alpha_1} \wedge \ldots \wedge d \theta^{\alpha_s} \wedge \delta^{(t_1)} \left( d \theta^{\beta_1} \right) \wedge \ldots \wedge \delta^{(t_q)} \left( d \theta^{\beta_q} \right) \,,\nonumber 
\end{eqnarray}
where $\delta^{(i)} \left( d \theta \right) \equiv \left( \iota \right)^i \delta \left( d \theta \right)$. The form number is obtained as 
\begin{equation}\label{ABRIFH}
	p = r + s - \sum_{i=1}^q t_i \ ,
\end{equation}
since the contractions carry negative form number. The two quantum numbers $p$ and $q$ in eq. \eqref{ABRIFH} correspond to the {\it form} number and the {\it picture} number, respectively, and they range as $-\infty < p < +\infty$ and $0 \leq q \leq m$, so the picture number counts the number of delta's. If $q=0$ we have superforms, if $q=m$ we have integral forms, if $0<q<m$ we have pseudoforms.

As in conventional geometry, we can define the integral of a top form on a supermanifold (more rigorously, the integration 
is on  the parity-shifted tangent space $\Pi T\mathcal{SM}$) as
 \begin{eqnarray}\label{ABRIFI}
I[\omega] = \int_{{\cal SM}} \omega_{top}^{(n|m)} =  \int \omega(x, \theta)[d^{n}x d^{m}\theta]\ ,
\end{eqnarray}
where we integrated over the odd variables $dx$ and over the even variables $d \theta$ 
to obtain an ordinary superspace integral over the variables $(x, 
\theta)$.  


\end{document}